# IDENTIFYING EXOPLANETS WITH MACHINE LEARNING METHODS: A PRELIMINARY STUDY

Yucheng Jin, Lanyi Yang and Chia-En Chiang

EECS Department, University of California-Berkeley, Berkeley, CA, USA

### *ABSTRACT*

*The discovery of habitable exoplanets has long been a heated topic in astronomy. Traditional methods for exoplanet identification include the wobble method, direct imaging, gravitational microlensing, etc., which not only require a considerable investment of manpower, time, and money, but also are limited by the performance of astronomical telescopes. In this study, we proposed the idea of using machine learning methods to identify exoplanets. We used the Kepler dataset collected by NASA from the Kepler Space Observatory to conduct supervised learning, which predicts the existence of exoplanet candidates as a three-categorical classification task, using decision tree, random forest, naïve Bayes, and neural network; we used another NASA dataset consisted of the confirmed exoplanets data to conduct unsupervised learning, which divides the confirmed exoplanets into different clusters, using k-means clustering. As a result, our models achieved accuracies of 99.06%, 92.11%, 88.50%, and 99.79%, respectively, in the supervised learning task and successfully obtained reasonable clusters in the unsupervised learning task.*

## 1. INTRODUCTION

During the past decades, astronomers have been relentlessly searching for habitable exoplanets (planets outside the Solar System) with great interest [1]. They have devoted a considerable amount of time, energy, and money to explore the vast universe with specialized equipment, such as the astronomical telescope. As a result, astronomers can collect exoplanet data for further analytics, but the identification of a potential exoplanet remains a complicated problem.

There are several traditional exoplanet identification techniques, including the wobble method, direct imaging, gravitational microlensing, etc. The wobble method detects exoplanets via a Doppler shift in the star's light frequencies caused by its planets; direct imaging uses extra-terrestrial telescopes to capture images from exoplanets; gravitational micro lensing analyses the distortion of the background light [2]. These traditional methods are expensive, time-consuming, and sensitive to variation in the measurement. Based on these limitations, our motivation for this study is to develop a simple, efficient, and economic approach to identify exoplanets. Therefore, we proposed the idea of using machine learning tools to facilitate the identification of exoplanets and used two datasets collected by NASA to conduct both supervised learning and unsupervised learning.

For the supervised learning part, we used the Kepler dataset consisted of false positive (labelled as -1), candidate (labelled as 0), and confirmed (labelled as 1) exoplanet samples to perform three-categorical classification. The objective is to achieve a classification accuracy of more than 95%.We selected and trained four models for this problem: classification tree, random forest, naïve Bayes, and neural network. As a result, our models achieved accuracies of 99.06%, 92.11%, 88.50%, and 99.79%, respectively.

For the unsupervised learning part, we used another dataset consisted of the confirmed exoplanets data to do clustering. The objective is to find exoplanets that are in the same cluster as the earth. We selected and trained a k-means model for this problem. As a result, our model partitioned reasonable clusters, and we visualized all exoplanets in the same cluster as the earth in a star map.

The rest of this paper is organized as follows. Section II summarizes the related work from other researchers. Before the modelling and inference, we conducted data cleaning, exploratory data analysis (EDA), feature selection, etc., and these are written in Section III. In Section IV, we present the detailed experimental setup, including model optimization and parameters. Section V analyses the results obtained by our models and discusses the meanings of the results. Finally, Section VI concludes the paper with the significance of this study and future work to be done.

## 2. RELATED WORK

NASA's Kepler Mission devoted years of effort to discover exoplanets. Natalie M. Batalha [3], stated that the progress made by the Kepler Mission indicates that there are orbiting systems that would be potentially habitable places for the human beings. Lissauer et al. [4] emphasized the contribution made by the Kepler Mission and the significance of its data in exoplanets detection and identification.

Based on the Kepler dataset or other similar datasets, there are studies that combine machine learning with exoplanets identification. Christopher J. Shallue and Andrew Vanderburg [5] used a convolutional neural network to process the astrophysical signals and validated two new exoplanets according to the CNN predictions. Rohan Saha [6] implemented logistic regression, decision tree, and neural network on the Kepler dataset to find out the probability of existence of an exoplanet candidate. Maldonado et al. [7] summarized similar work in exoplanet transit discovery with ML-based algorithms. These related studies provide us with precious insights into the data pre-processing, model selection, training, and inference procedures.

## 3. DATA PRE-PROCESSING

In this section, we introduce the data pre-processing procedures involved in this study and a short description of the datasets.

### 3.1. Dataset Description

The first dataset (the Kepler dataset) in this study is collected by NASA from the Kepler Space Observatory. In 2009, NASA launched the Kepler Mission, an effort to discover exoplanets with the goal of finding potentially habitable places for human beings [8]. The mission lasted for over nine years with remarkable legacies—a total number of 9,564 potential exoplanets are contained in the Kepler dataset, each associated with features that indicate the characteristics of the exoplanet. Among these features, there is a categorical target variable, koi_disposition, with three possible values, "CONFIRMED" (labelled as 1), "CANDIDATE" (labelled as 0), and "FALSE POSITIVE" (labelled as -1). If an exoplanet is "CONFIRMED", we know its existence has been confirmed, and is associated with a name recorded by kepler_name variable; if an exoplanet is "CANDIDATE", its existence has not been proven yet; if an exoplanet is "FALSE POSITIVE", it has been proven a negative observation. There are totally 2,358 confirmed exoplanets, 2,366 candidate exoplanets, and the rest 4,840 exoplanets are false positive. We used the Kepler dataset for supervised classification.

The second dataset (the confirmed exoplanets dataset) contains stellar and planetary parameters of the confirmed exoplanets [9]. Crucial parameters include radius, mass, density, temperature, etc. There are totally 4,375 confirmed exoplanets in this dataset, some are discovered by the Kepler Space Observatory, the rest are found by other space observatories. We used the confirmed exoplanets dataset for unsupervised clustering.

## 3.2. Data Cleaning

The first step of data cleaning was to calculate the proportion of empty entries in each column. We excluded columns with a large proportion of empty data. Then columns with a small fraction of empty data were manually selected and the empty entries in the preserved columns were filled with proper values. Finally, some data were dropped if they had empty values after data cleaning.

## 3.3. Exploratory Data Analysis (EDA)

Before feature selection and model construction, exploratory data analysis (EDA) is necessary to facilitate the data analysis process and make it easier and more precise [10]. In this study, EDA was carried out to obtain an intuitive high-level understanding of the datasets.

### 3.3.1. Correlation Analysis

We first conducted correlation analysis to identify highly correlated variables to reduce data redundancy and collinearity. The correlation matrices of the two datasets are shown in Fig. 1 and Fig. 2.

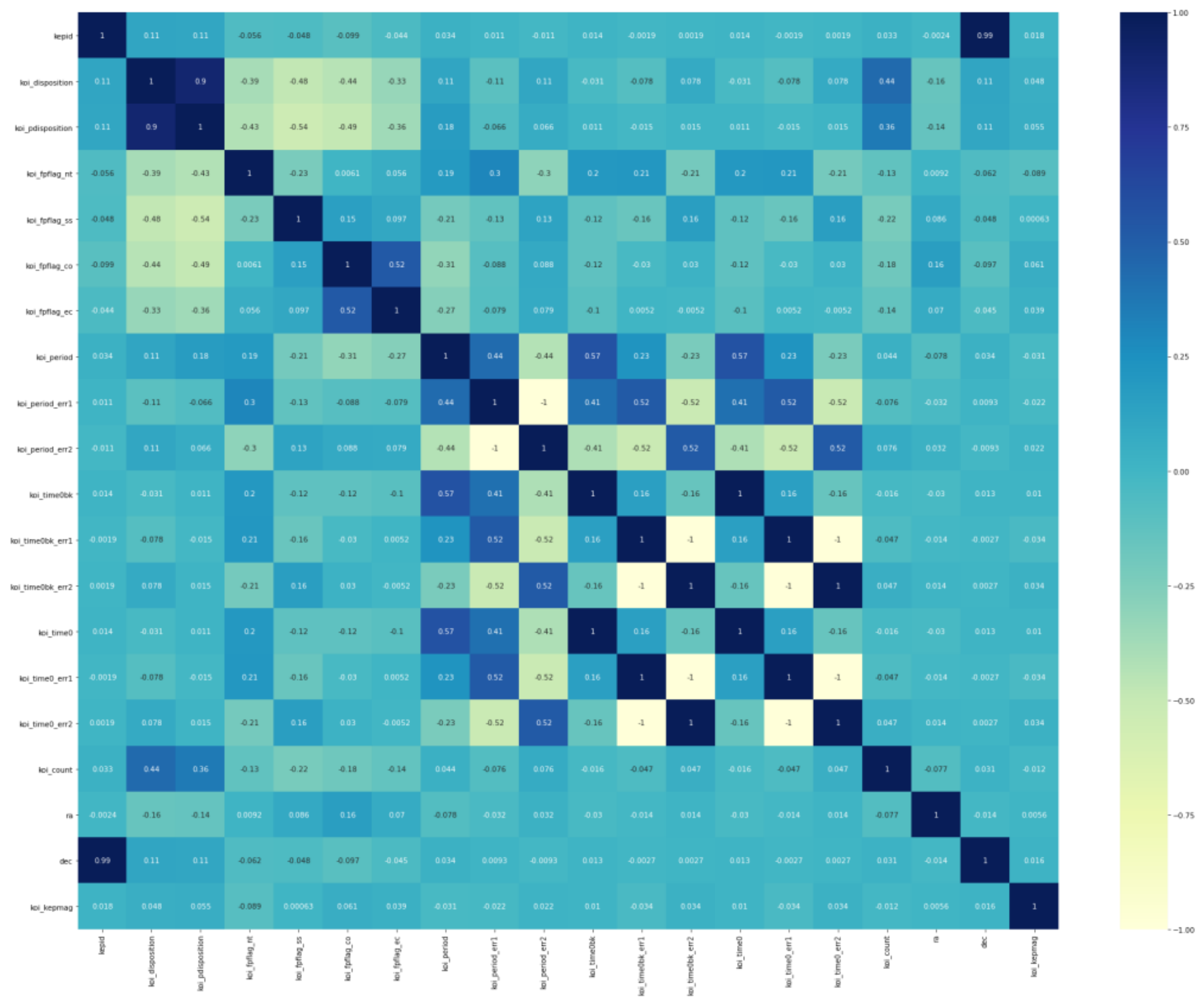

Figure 1. Correlation Matrix of the Kepler Dataset

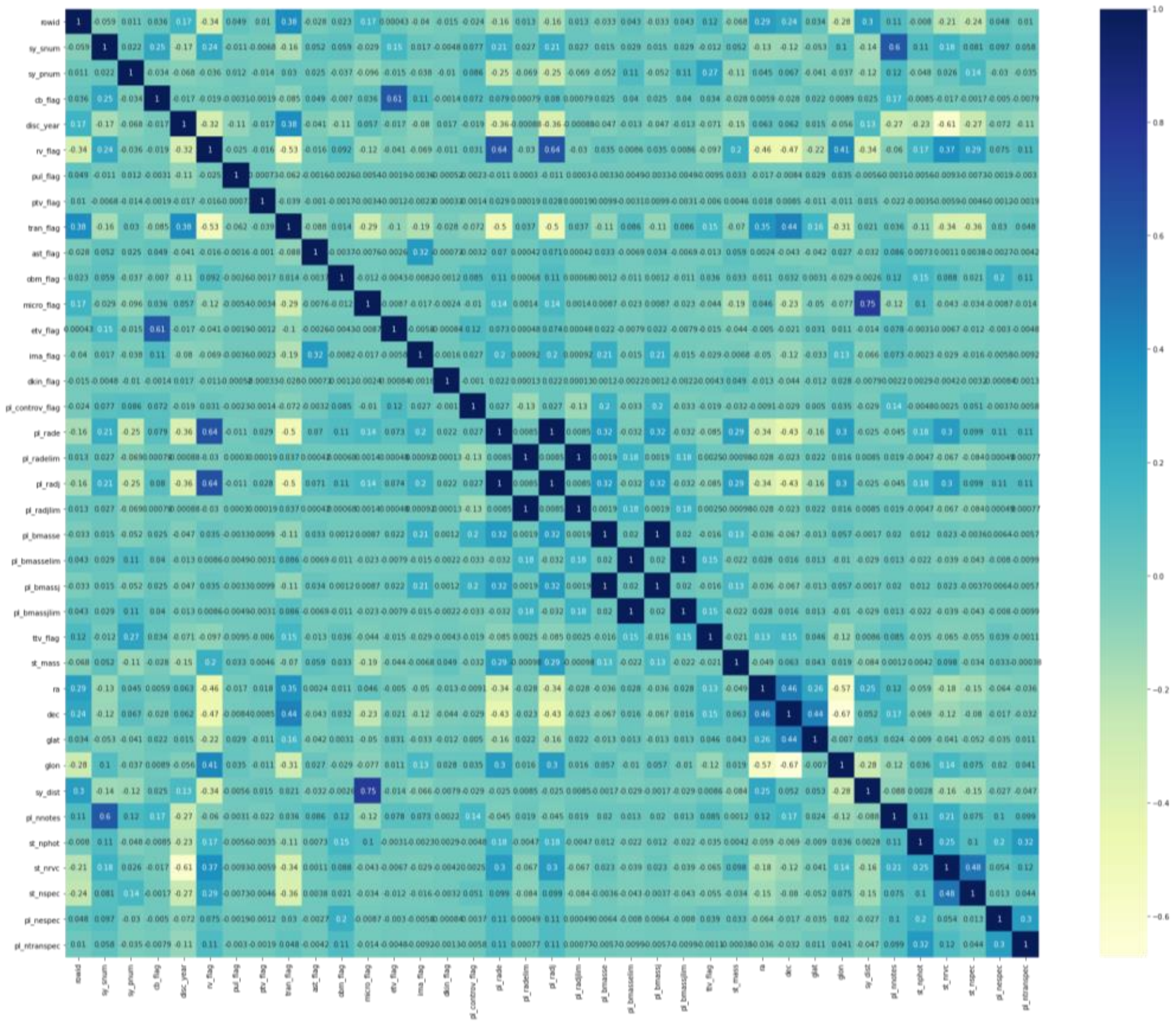

Figure 2. Correlation Matrix of the Confirmed Exoplanets Dataset

### 3.3.2. Univariate Analysis

Then univariate analysis was then performed to visualize the distribution of each variable.

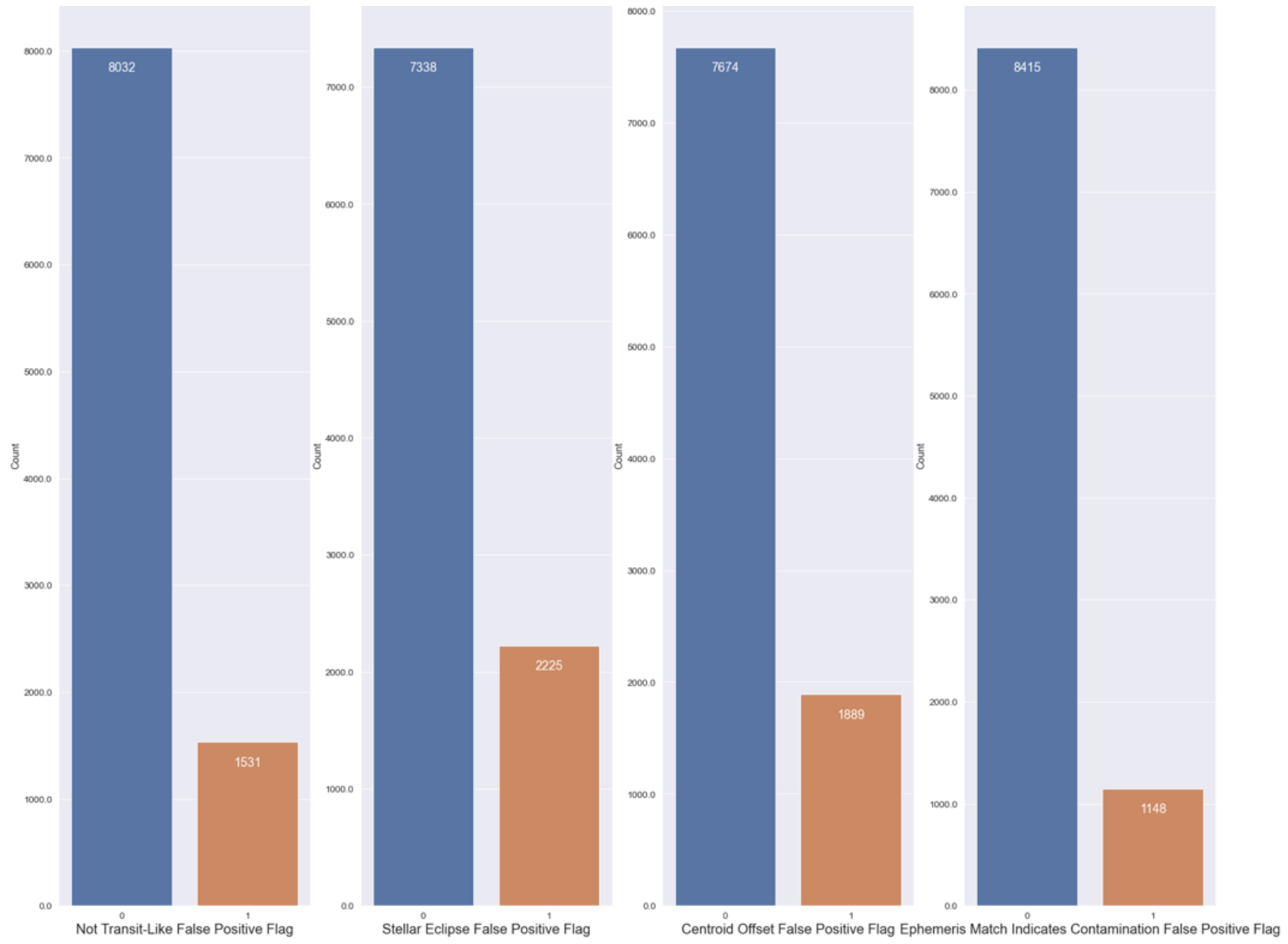

Figure 3. Count Plots of the Binary Variables of the Kepler Dataset

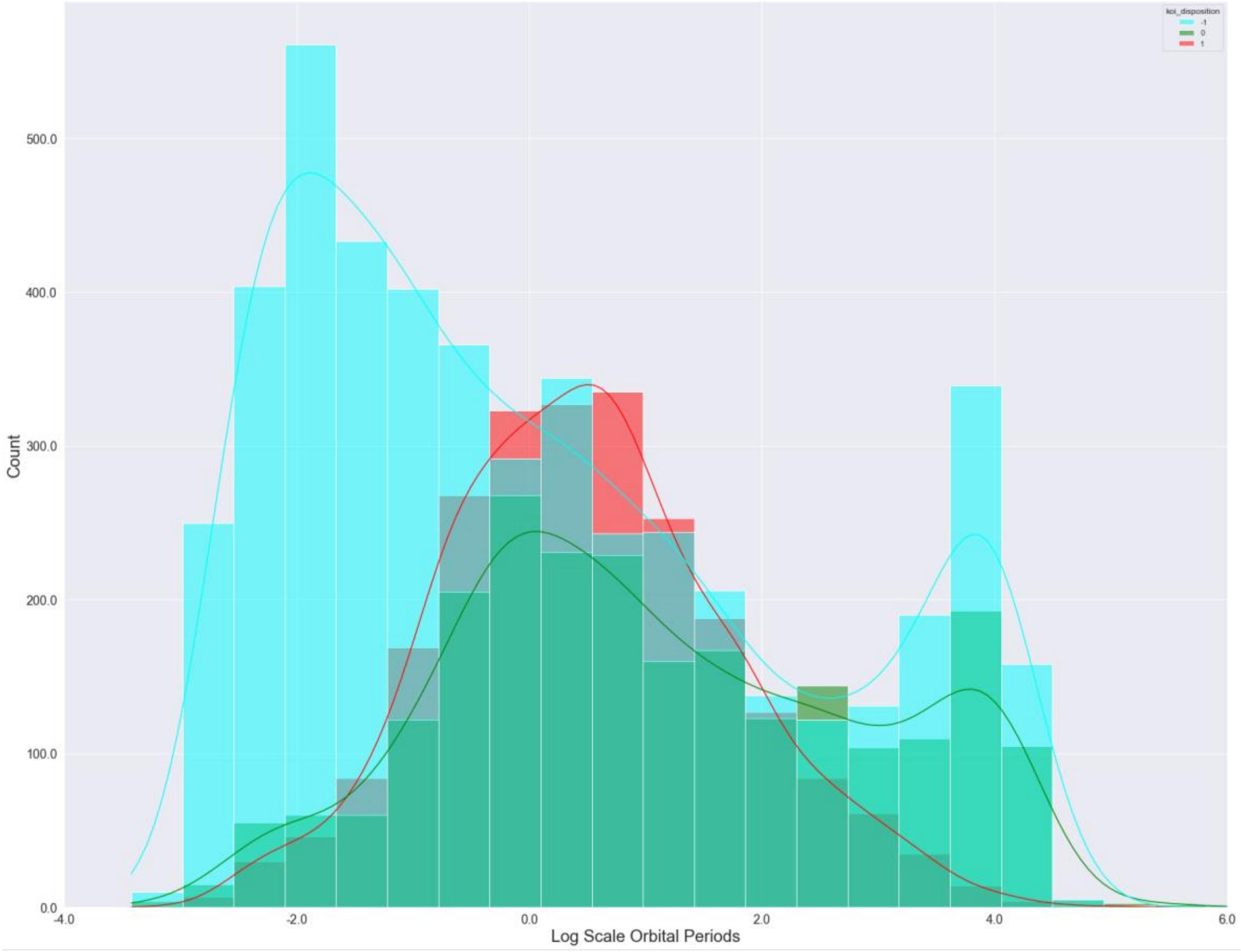

Figure 4. Histogram of the Log Scale Orbital Periods of the Kepler Dataset

Fig. 3 shows the count of each binary variable in the Kepler dataset, which gives the distribution of four binary characteristics (non-transit-like, stellar eclipse, centroid offset, ephemeris match indicates contamination) of the exoplanet candidates.

Fig. 4. is a histogram that indicates the relationship between an important feature, the log scale orbital period, and the target variable. From Fig. 4, the confirmed exoplanets follow a Gaussian distribution with respect to the log scale orbital period in the range between -2 and 4. If the value of the log scale orbital period is too high or too low, the candidate is more likely to be a false positive observation.

### 3.3.3. Bivariate Analysis

Finally, we used bivariate analysis to investigate the pairwise relationship of different features and observe how they affect the target variable.

Fig. 5 is a star map of the exoplanet candidates. In this star map, each pair of celestial coordinates is measured by right ascension (RA), the celestial coordinate that represents longitude, and declination (Dec), the celestial coordinate that represents latitude [11]. Fig. 5 demonstrates that there is no strong correlation between the target variable and coordinates.

Fig. 6 is a categorical plot of the log scale orbital period versus number of planets in the exoplanet candidate's solar system. The result shows that false positive samples are most likely to have just one or two planets in their solar system. With a higher number of planets in a candidate's solar system, the probability of it being a real exoplanet is also higher.

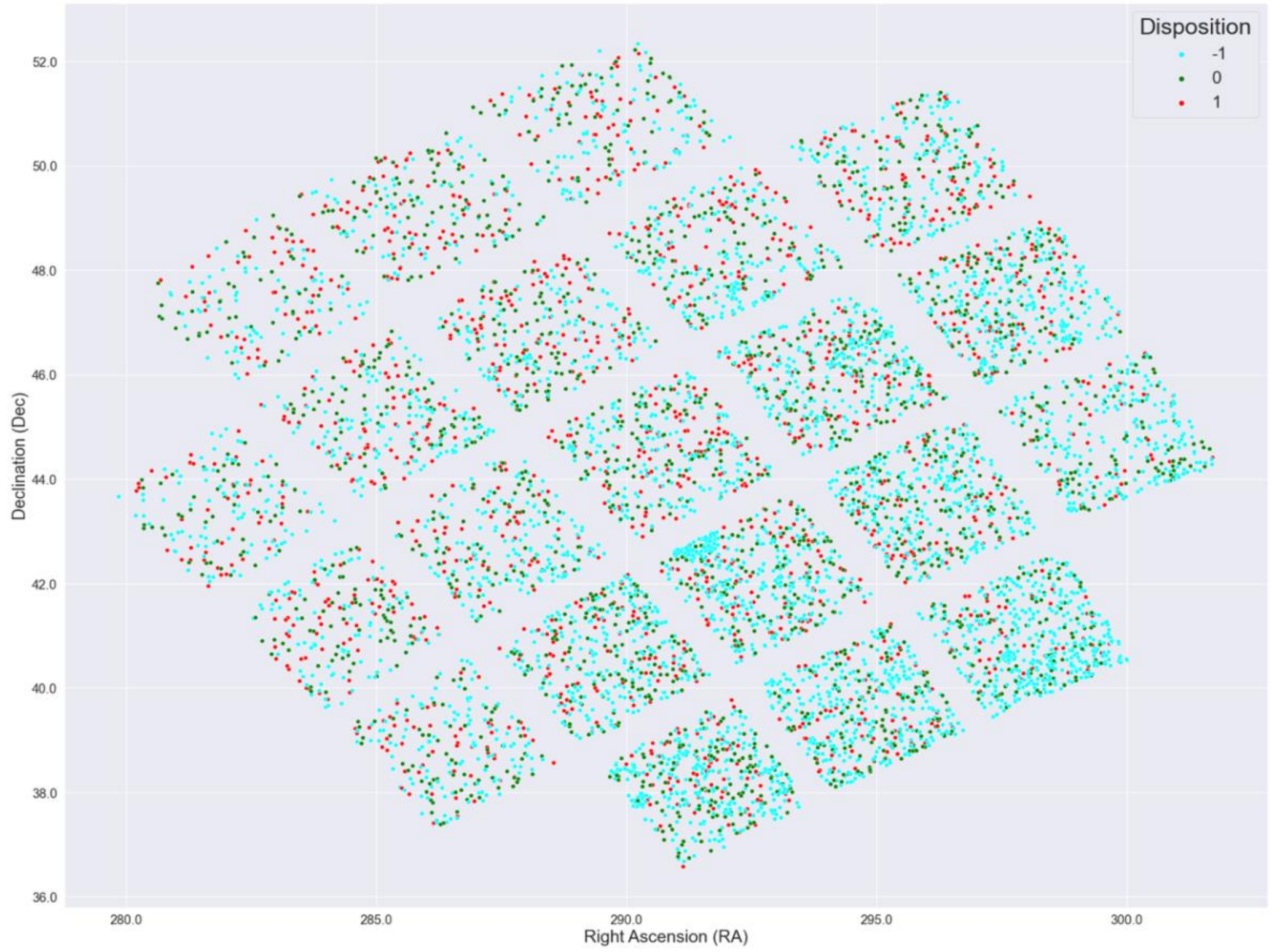

Figure 5. Star Map by Right Ascension (RA) and Declination (Dec)

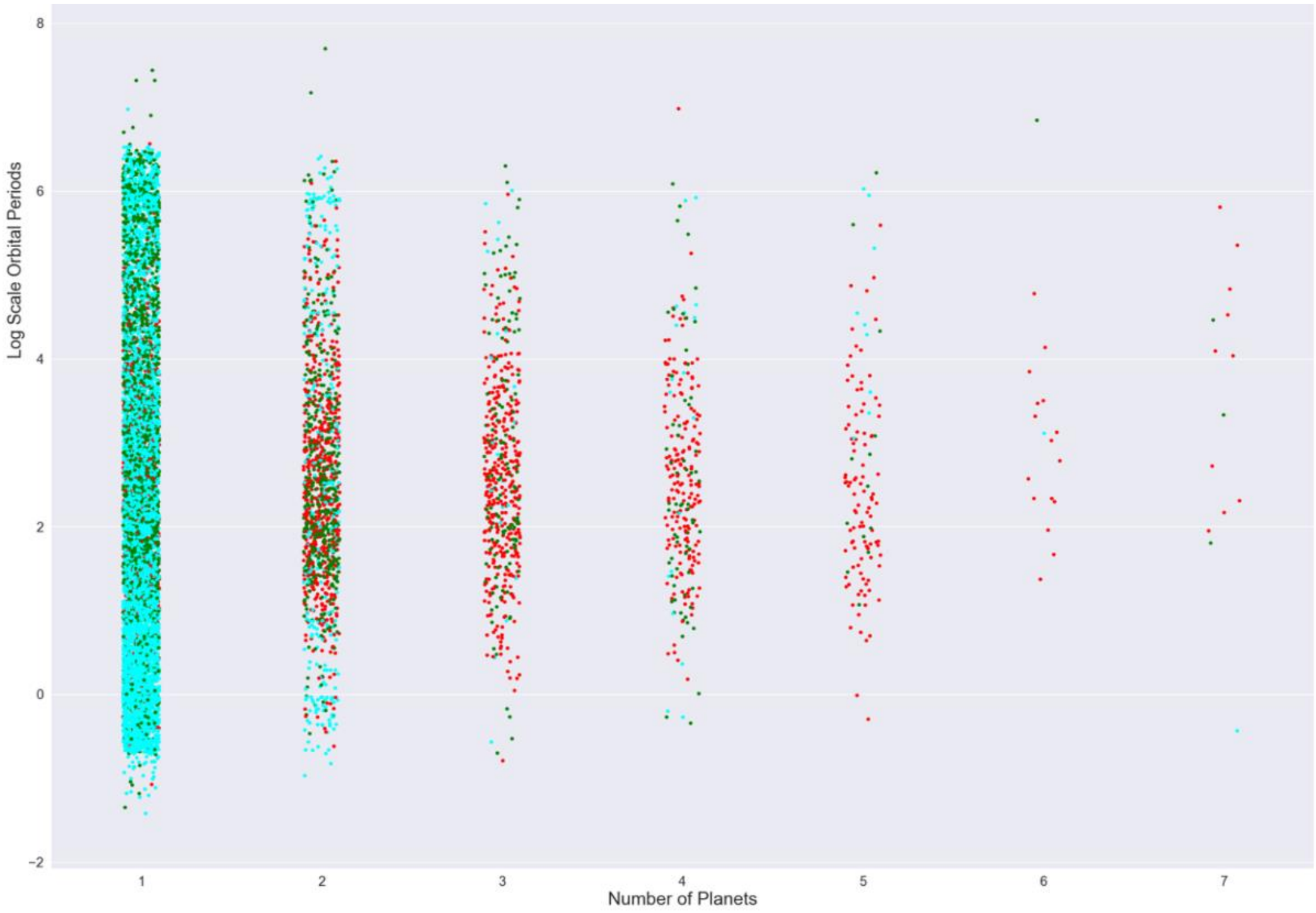

Figure 6. Categorical Plot of the Log Scale Orbital Period versus Number of Planets

### 3.4. Feature Selection

Based on the correlation heatmaps, we excluded variables that have high correlation coefficients with one or more other input features or have a low correlation coefficient with the target variable. For example, variables that store the deviation of observations, such as koi_period_err (error of the orbital period) and koi_time0_err (error of the transit epoch), are highly dependent on the observed values (orbital period and transit epoch). Therefore, we just kept variables that store the observed values but abandoned variables that record deviation. Some preserved features are listed in Table 1.

Table 1. Examples of Preserved Features after Feature Selection

| Variable | Dataset | Meaning | Description |
|---|---|---|---|
| kepler_name | Kepler | Kepler Name | Kepler name in the form of “Kepler-” plus a number and a lower-case letter (e. g. Kepler-22b, Kepler-186f) |
| koi_fpflag_nt | Kepler | Non-Transit-Like Flag | 1 means the light curve is not consistent with that of a transiting planet |
| koi_period | Kepler | Orbital Period | Measured in days and is taken in log scale |
| koi_count | Kepler | Number of Planets | Number of exoplanet candidates identified in a solar system |
| pl_radj | Confirmed Exoplanets | Planet Radius | Measured in Jupiter Radius |
| pl_bmassj | Confirmed Exoplanets | Planet Mass | Measured in Jupiter Mass |
| sy_dist | Confirmed Exoplanets | Distance | Distance to the planetary system in parsecs |

## 4. EXPERIMENTAL SETUP

In this section, we discuss the model training process in detail with model optimization and model parameters. The training process was divided into two parts, training models for classification, and training models for clustering. For these two parts, we used the Kepler dataset and confirmed exoplanets dataset, respectively. For the supervised learning part, we used decision tree, random forest, naïve Bayes, and neural network. For the unsupervised learning part, we used k-means clustering.

### 4.1. Classification Tree

The reason why we chose the decision tree model is that it can split data with classification rules based on the highest information gain. In a classification tree model, each internal node represents a sample to be split, and each leaf node represents a class label, or prediction. We implemented the classification tree model using DecisionTreeClassifier from scikit-learn library and optimized its performance with different parameters. We set the minimum number of samples required to split an internal node from 2 to 100 and maximum depth of the tree from 1 to 20, then we selected the parameters with the highest accuracy. As a result, the optimal minimum number of samples required to split an internal node is 53 and maximum depth of the tree is 7.

### 4.2. Random Forest

Random forest combines multiple decision trees at the training time. We implemented the random forest model using RandomForestRegressor from scikit-learn library. We optimized the number of trees in the forest by setting this parameter from 20 to 1000 with a step size of 20. As a result, the highest accuracy was obtained when the number of trees is 40.

### 4.3. Naïve Bayes

Naïve Bayes applies Bayes' theorem under the assumption that features are independent. It's a simple probabilistic approach to conduct the classification task. We implemented the naïve Bayes model using ComplementNB from scikit-learn library. Before training the naïve Bayes classifier, we standardized input features, encoded the target variable as a one-hot vector, and set each prior as the proportion of each class in the sample space.

### 4.4. Multilayer Perceptron

Neural network is another machine learning model that is widely used in classification problems. We implemented the neural network model using MLP Regressor from scikit-learn library. We tried three activation functions, logistic, tanh, and ReLU. These activation functions determine how the weighted sum of the input to the current layer is converted into the output to the next layer.We also chose three types of solvers, stochastic gradient descent, quasi-Newton method, and Adam optimizer. Finally, we optimized the layer size and learning rate. The optimal neural network uses tanh as the activation function and Adam as the optimizer with a layer size of 25 and a learning rate of 0.003.

### 4.5. K-means Clustering

K-means is a classic unsupervised learning algorithm for clustering problems. It partitions data into k clusters with the objective of minimizing the sum of distance of each sample to its cluster centroid in a repeated way. We aimed at finding out which exoplanets are in the same group as the earth that might indicate potentially habitable places for human beings. We implemented the k-means model using KMeans from scikit-learn library.

## 5. Experimental Results and Analysis

For the classification problem, the decision tree achieved an accuracy of 99.06%, random forest achieved an accuracy of 92.11%, naïve Bayes achieved an accuracy of 88.50%, and multilayer perceptron achieved an accuracy of 99.79%. These four models all performed well on the test set.

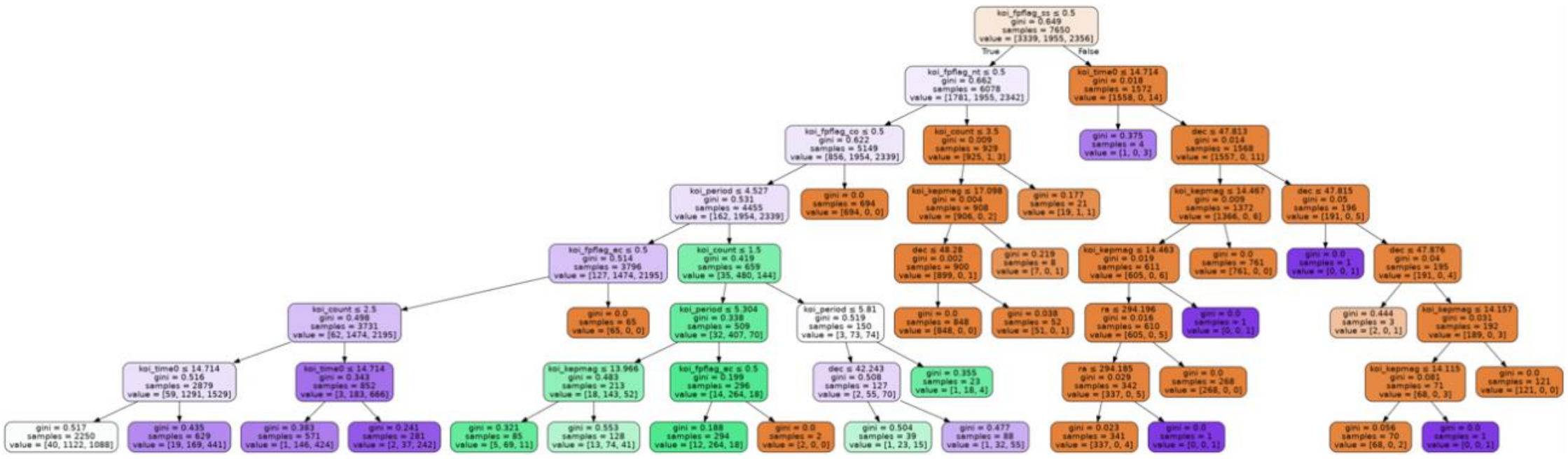

Figure 7. The Optimal Classification Tree Obtained

We visualized the classification tree with the highest accuracy using graphviz and plotted the importance of each feature. The optimal classification tree obtained is shown in Fig. 7 and the importance of each feature is shown in Fig. 8 and Table 2.

We further evaluated the performance of these models with 10-fold cross-validation using KFold from scikit-learn library. As a result, the random forest model achieved the best average accuracy of 82.39%.

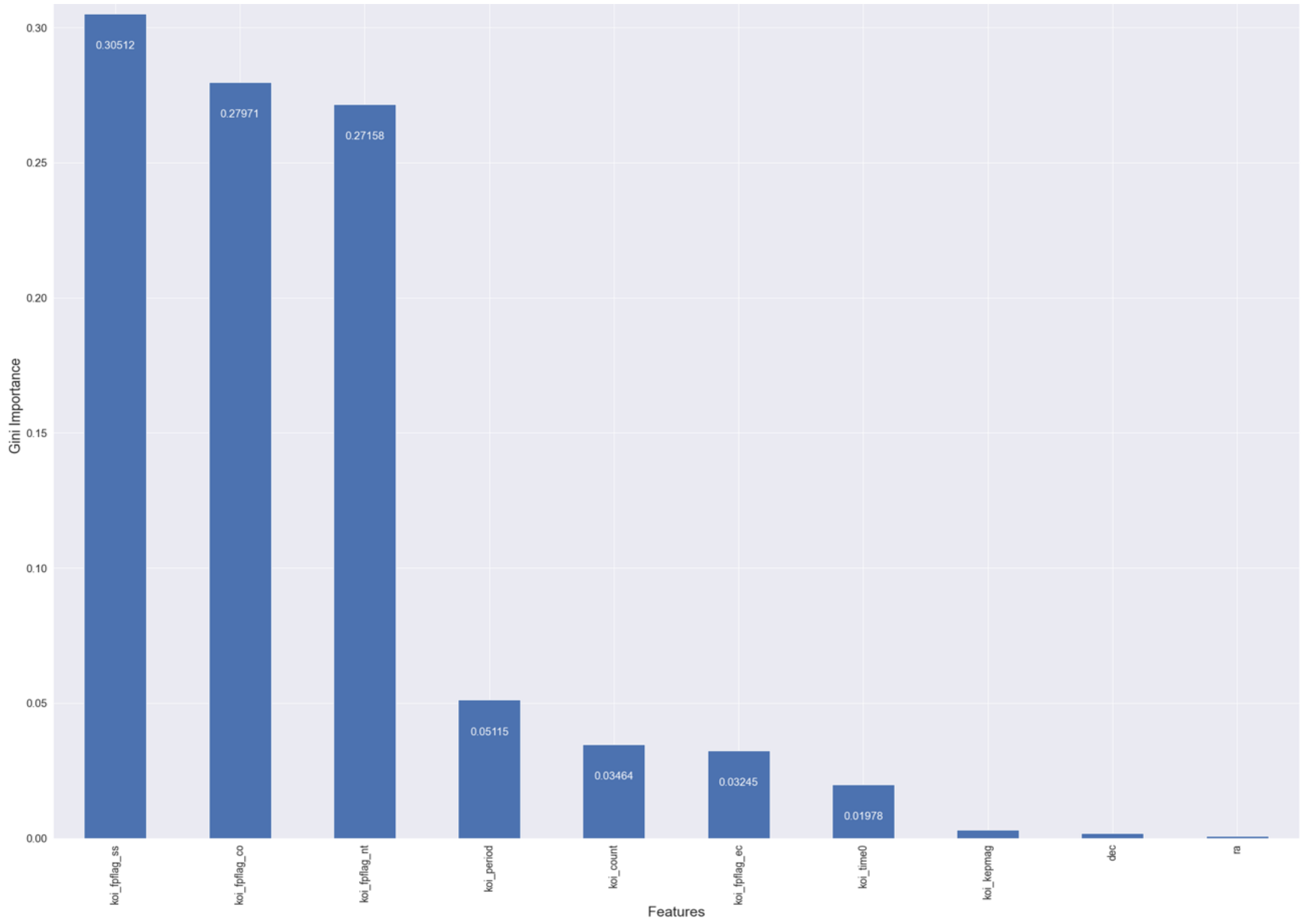

Figure 8. The Importance Scores of Features in the Kepler Dataset

Table 2. Kepler Feature List with Variable Name, Importance, Meaning, and Description

| Variable | Importance | Meaning | Description |
|---|---|---|---|
| koi_fpflag_ss | 0.30512 | Stellar Eclipse | 1 means some phenomena caused by an eclipsing binary observed |
| koi_fpflag_co | 0.27971 | Centroid Offset | 1 means the source of the signal is from a nearby star |
| koi_fpflag_nt | 0.27158 | Non-Transit-Like | 1 means the light curve is not consistent with that of a transiting planet |
| koi_period | 0.05115 | Orbital Period | Measured in days and is taken in log scale |
| koi_count | 0.03464 | Number of Planets | Number of exoplanet candidates identified in a solar system |
| koi_fpflag_ec | 0.03245 | Ephemeris Match Indicates Contamination | 1 means the candidate shares the same period and epoch as another object |
| koi_time0 | 0.01978 | Transit Epoch | Measured in Barycentric Julian Day (BJD) |

Finally, we visualized the exoplanets in the same cluster as the earth in Fig. 9. As a result, these exoplanets are most likely to be habitable for human beings.

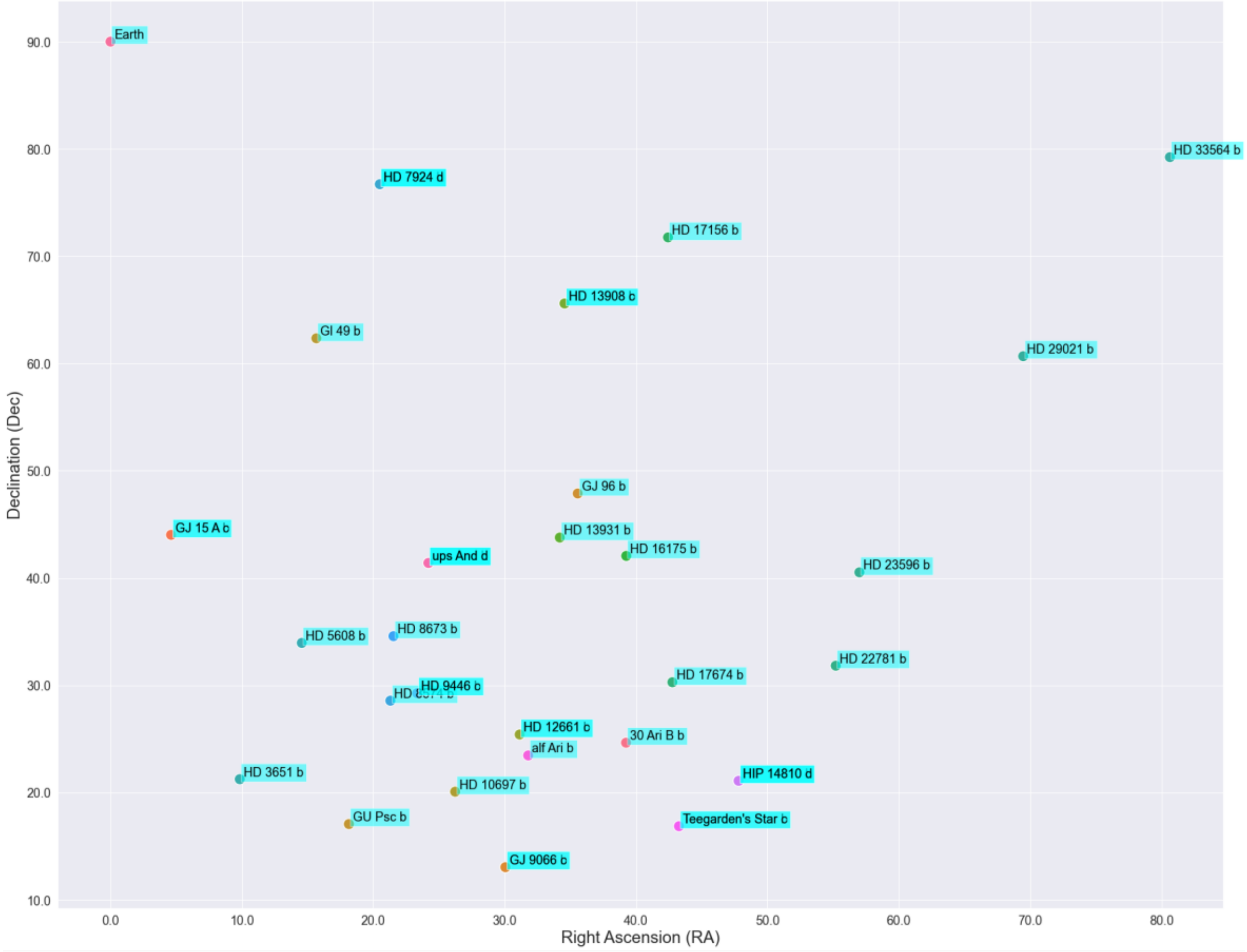

Figure 9. Star Map of the Exoplanets in the Same Cluster as the Earth

## 6. CONCLUSION

In this project, we conducted both supervised and unsupervised learning on two datasets collected by NASA, the Kepler dataset and confirmed exoplanets dataset. The Kepler dataset was used to predict the existence of exoplanet candidates by classification techniques, including decision tree, random forest, naïve Bayes, and neural network. The confirmed exoplanets dataset was used to find habitable exoplanets by picking up exoplanets in the same cluster as the earth.

For the supervised learning task, comprehensive data cleaning and EDA were performed to help visualize the data and get a higher-level understanding of the samples. Then correlation-based feature selection was conducted, several redundant features were removed because of low feature-target correlation or high feature-feature correlation. The selected machine learning models were then implemented. The optimal hyper-parameters were obtained by experiment and accuracy was improved. Our models finally achieved accuracies of 99.06%, 92.11%, 88.50%, and 99.79%, respectively, and were compared by 10-fold cross validation. As a result, the random forest model performed the best among these four.
For the unsupervised learning task, basic EDA and feature selection were also performed. Then the earth's features were added to the dataset before clustering. All exoplanets were divided into 100 clusters and exoplanets that are most likely to be habitable were found in the cluster that contains the earth.

As an extension of this project, it might be interesting to create new features from the current feature set using feature engineering techniques, because they can be helpful to improve the model performance. In addition, we can investigate the receiver operating characteristics (ROC) and the precision-recall curves to understand the diagnostic ability of different machine learning models. McNamara's test can also be applied to compare different algorithms [6]. Finally, the discussion of statistical significance tests and execution time can be included in model evaluation.

People from different cultures may connect to the concept of a "planet" as follows, they live on one, our mother earth, view the moon shared by all human beings, and learn the names of the other planets in our solar system from an early age. Planets that circle the star, rather than nebulae or galaxies, are easier to fit into our shared cultural view of the universe. For us, exoplanet exploration bridges the heaven with human consciousness and opens a vast exploration area to look forward to—seeking other habitable worlds. Finally, it has increased the likelihood that our long-term study points toward, we are not alone in the universe.

## AUTHORS

**Yucheng Jin** is a graduate student at the EECS department of University of California, Berkeley. His research interests are engineering applications of machine learning, machine learning security, and brain-machine interface.

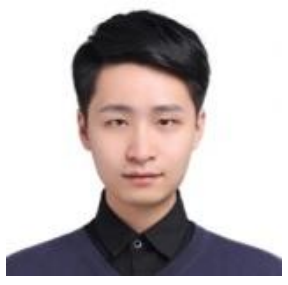

**Lanyi Yang** is a graduate student at the EECS department of University of California, Berkeley.

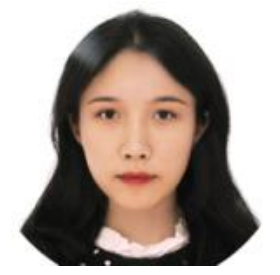

**Chia-En Chiang** is a graduate student at the EECS department of University of California, Berkeley.

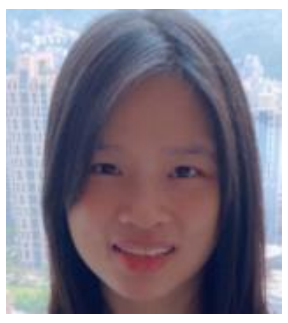